# ON THE CHARACTERISTICS OF THE FREE MARKET IN A COOPERATIVE SOCIETY

Norbert N. Agbeko

Email: truefreemarket@icloud.com

***ABSTRACT***

The key characteristic of a true free market economy is that exchanges are entirely voluntary. When there is a monopoly in the creation of currency as we have in today's markets, you no longer have a true free market. Features of the current economic system such as central banking and taxation would be nonexistent in a free market. This paper examines how currency monopoly leads to the instabilities and imbalances that we see in today's economy. It also proposes that currencies should emerge from the voluntary exchange of goods and services, and studies economic interaction across all scales, by considering economic action in cases where the self-interests of individuals are coincident. By examining the voluntary exchange of goods and services at the scale of an entire society, it is shown that a new currency system, which resolves a lot of the problems caused by the current fiat currency system, emerges naturally from the free market. The new currency system is robust and efficient, and provides a way for public goods and services to be provided, and its providers compensated, without the need for direct taxation. *JEL Codes:* E14, E42.

## I. INTRODUCTION

In our world today, currency arises from banking. Currency is initially created by a central bank and then expanded by banks through fractional reserve banking. The question of what the best way is for currency to be created is an open one. Our currency is generated from banking because of the way it evolved. We argue in this paper that this kind of currency is responsible for a lot of the imbalances and instability in today's economy. Currency should instead be generated from the voluntary contracts entered into by people within the free market.

The financial crisis of 2008 and the resultant global economic crisis generated an increased amount of scrutiny on the world's financial and currency systems and their role in the crisis. More people are beginning to suspect that the instabilities and imbalances in the market are a consequence of the currency system, and alternative currencies are increasing in popularity. It has long been held by Austrian economists that the recurring boom and bust phenomenon is not a characteristic of free markets, as suggested by Marx (Rothbard, 1996), but rather a consequence of the banking system (Mises, 1996). More precisely, the boom and bust cycle is a consequence of excessive credit expansion by the banking sector, supported by artificially low interest rates set by a central currency-issuing authority such as a central bank. Additionally, the inflationary effect of the method of currency creation systematically transfers purchasing power from the poorest in the society, through the wealthiest, to the banking and financial sector. The redistributive effect of inflation, and by extension the currency system, can be seen in the ever widening wealth gap between the rich and the poor, and the increasing concentration of wealth in the financial and banking sector.

The free market has been appropriated a lot of the blame for the recent economic crisis by some economists and politicians. They call for solutions to be implemented to control the excesses of the free market—which they blame for the crisis—arguing that free markets do not work properly without strong and effective regulation by the government. However, if one were to ask a free market economist if we actually have a free market, the answer would almost certainly be in the negative. Even those who would attribute the recent economic woes to the free market will usually admit that we do not have a true free market, although they seem to ignore the fact that their blame is consequentially misdirected.

Several reasons could be given for why the current system does not constitute a free market, but the most significant are legal tender laws and the existence of central banking, which are attributes of a monopoly in the creation of currencies. The argument presented by this paper is that currency monopoly is the root of the imbalances and instabilities in the world's economies. A core aspect of currency monopoly is the power that has been granted to governments to tax the general population within their respective countries. The ability to tax is the primary source of the currency monopoly because even in a society with competing currencies, a government must choose one or few currencies in which taxes will be paid. Such currencies become preferred over others because everyone engaging in economic activity has to pay taxes. People would simply prefer to transact in the currency or currencies used for the settlement of taxes because they would otherwise be faced with the overhead of converting other currencies to the ones preferred by the government.

To properly address the instabilities and imbalances in the world economy, it would be essential to eliminate taxation and the monopoly of currency creation held by governments in partnership with the financial and banking sector. We must do much more than merely look for ways to fix the current system or switching to previously used currency systems or variations thereof. A complete reexamination of the nature of money and credit together with an understanding of the causes of currency monopoly is needed.

The central thesis of this paper is that with the advent of civilization, the world has never had a true free market. This is because a true free market would not include governments having the power to tax the citizenry, or having monopolistic control over the creation of currency. This raises the obvious question: What does a free market actually look like? While this paper will not completely answer that question, it will examine the free exchange of goods and services between two parties in the context of a cooperative society and show that a new currency system—which addresses the problems with the current one—emerges naturally from the free market, and a picture of how free markets work in a civilized society starts to form. The work presented in this paper is a reformulation of the concepts of money and credit, and an extension of the study of free markets to include economic interaction when the self-interested actions of individuals in a society are coincident, resulting in mutual interests. The meanings of some terms commonly used by economists have been changed in this paper for the sake of greater precision and readers are asked to bear this in mind.

## II. BACKGROUND

Money is most commonly defined as a medium of exchange. While this definition is technically correct, its vagueness in meaning allows for a variety of things to be considered as kinds of money and hence complicates the analysis of the nature of money. A theory of the origin of what we call money today was described by Menger (2004). In summary, the idea is that people would start off by exchanging goods directly, otherwise known as barter. However, depending on the marketability of the goods or services produced by the individual actors in the economy, they would experience varying degrees of difficulty in exchanging their goods or services for what they actually want. An individual may have to exchange his product for a more marketable good first, and then exchange the more marketable good for the product that he wanted in the first place. This person may actually have to exchange goods several times before he obtains a good that he can exchange for what he originally wanted. Clearly it would be desirable to minimize the number of intermediate exchanges, and this can be done by simply acquiring the most marketable of goods. Over time, certain commodities emerge as the most marketable and these become the media of exchange which we today call money.

The objection to this theory of the origin of money is not about the process but rather the idea that the highly marketable commodities should be considered as media of exchange. It would be more precise to say that they emerge as a medium of indirect exchange since that is the process they facilitate. The question then is: should money be defined as a medium of indirect exchange? This would be an incomplete definition since it doesn't account for the fact that the money is part of the exchange and does not merely facilitate it. The kind of money that arises from Menger's theory of the origin of money is what Mises (1953, pp. 59–62) calls commodity money. It can at least partially be thought of as a medium of indirect exchange. In addition, since the highly marketable commodities are not exchanged for their usage value, there is a clear distinction between seller and buyer in the exchange. The buyer exchanges the highly marketable commodity for the good or service that he is going to consume. In this paper, the terms money and commodity money are considered equivalent, and are defined as a medium of payment rather than a medium of exchange.

## III. MEDIUM OF EXCHANGE

It should be noted that any exchange is in two parts. If two parties engage in an exchange, you have a situation where one party provides a good or service to a second party, and then at some point later in time, the second party compensates the first by providing a good or service in return. In a barter exchange the time interval between the two parts of the exchange is very small and so no explicit medium of exchange is required. However, if there is a significant time interval between the two parts, a form of contract is required. The party receiving the good or service in the first part of the exchange must honor this contract when it is redeemed, by providing the good or service specified in the contract to its bearer. In other words, the receiver of the good or service in the first part of the exchange incurs a debt to the provider, or equivalently the provider in the first part of the exchange earns a credit against the receiver of the good or service. The exchange is complete when the contract is honored.

In *The Theory of Money and Credit*, Mises (1953, pp. 59–62) defines three kinds of money—namely commodity money, credit money and fiat money. By the definition of money used here only commodity money is money. Credit money in this paper will simply be referred to as credit. While money was defined as a medium of payment, credit will be defined as a medium of exchange. This definition of credit as a medium of exchange is based on the fact that each exchange has two parts, which are separated in time and possibly also in space. The medium of exchange is simply the contract, whether implicit or explicit, that links the two parts of the exchange. The reason for defining money—which is a commodity—as a medium of payment is now clear: money is part of the exchange and does not simply link the two parts of the exchange as credit does. With this definition, the medium of exchange is not a physical good or service but a contract. The medium of exchange has no value in and of itself, but only has value with respect to the goods or services for which it can be redeemed. Finally, the medium of exchange is the claim that the party that provides the good or service in the first half of the exchange has to the labor of the other party.

Credit is therefore a claim to goods and services, and in this framework would take the form of an IOU. It is easy to see that the contract that links the two parts of the exchange can be either in the form of an IOU or in the form of an invoice[1]. The IOU represents a credit to the bearer while the invoice represents a debt owed to the bearer. Both credit and debt are considered to be media of exchange. Credit and debt—and correspondingly IOUs and invoices—are defined to be claims to goods and services arising from voluntary exchanges in the free market. The only difference between them is which of the two parties in the exchange issues the contract. Credit, or the IOU, is issued by the receiver of the good or service in the first part of the exchange, while debt, or the invoice, is issued by the provider of the good or service in the first part of the exchange. For the remainder of this paper the terms IOU and invoice will be used instead of credit and debt respectively since they are more descriptive. IOUs and invoices may be implicit or explicit, and may be verbal or written. They have a finite lifetime, i.e., they are created during the first part of the exchange and terminated at the second part, which is the completion of the exchange.

---

[1] An invoice can be thought of as a you-owe-me or taking some grammatical license, a UOI, i.e., the inverse of an IOU.

## IV. CURRENCY

Currency is defined as any widely accepted medium of exchange or medium of payment. It is important to note that while a medium of payment or exchange is generated by economic interaction between two parties, it can only become currency in the context of an entire society. The first kind of currency that emerges in a society comes from money. The most marketable of goods emerge as currency if they can be denominated. For example, gold and silver are easily valued by their weight and so they emerged as currency in the form of coins. Carrying around gold, silver or other such marketable goods can be inconvenient and also pose a security risk, hence the need for banking arises when you have such a currency system. Banks originally started as a way to safely store one's gold or silver while transacting with certificates issued by the bank. These certificates were redeemable for the gold or silver stored in the bank vaults and are the origin of modern currency.

### IV.A. Money and Banking

The problem with money is that as it becomes currency, it makes sense to have some on hand, as this simplifies trading. However, two issues arise, namely security and portability. These issues led to the invention of banking, whereby a bank would provide storage for people's money and provide them with more portable and less risky banknotes. The banknotes could always be redeemed for the commodity.

In most places where banking was invented in this way, gold and silver were the primary commodities that became money. Banks would therefore hold people's gold and silver and give them banknotes. It is easy to see how such a system would naturally evolve into what we have today. There are two contributory factors, the first being the banks themselves. The use of banknotes as currency resulted in greater efficiency of the economy and hence increased economic growth. The banks would have noticed that there was an increasing demand for their services and their banknotes. But at the same time, people would typically not redeem the banknotes for the commodity held by the banks. The banks would have realized that they did not have to hold gold or silver in the exact amount of banknotes they had issued. They could issue more banknotes than the actual gold or silver that they held in reserve. This gave birth to what we call fractional reserve banking today. The reserve ratio—the ratio of the value of the banknotes issued to the value of actual gold or silver held by the bank—changed from one-to-one to many-to-one. Of course when the reserve ratio became too high, people would start to notice and be concerned about whether they could actually get their money back. This resulted in bank panics, where a large number of people suddenly wanted to redeem their certificates for their real money but the banks simply did not have enough in reserve.

The second factor was the government or state. The currency system that arose from banking, while unstable with its bank panics, could not give rise to what we have today without government. The reason for this is that governments collect taxes, purportedly because that is the only way for people to contribute towards the provision of public services. With banknotes becoming the currencies of choice, the government would want to start collecting taxes in banknotes. But it is clear that it is more convenient for the government if all banknotes were denominated in the same way. That way they would all be of equal value, could be redeemed at any bank, and would make the collection of taxes more straightforward. A partnership thus arose between the banks and the government, whereby the banking sector became responsible for the creation of currency.

With banknotes as the currency used for the settlement of taxes, all members of the citizenry must now at some point make use of this currency, as they have to pay taxes. This has the effect of eliminating currency choice as it makes sense to transact in banknotes rather than a different currency, because then you would not have to purchase the banknotes at asking price when you have to pay taxes. Any alternative currency that may evolve in this environment is automatically at a disadvantage. Finally, legal tender laws make it compulsory to accept banknotes—and also bank credit—as payment in the settlement of any debt. With the monetary monopoly now created, the move towards central banking and zero-reserve banking is straightforward. Currency is no longer generated by economic activity but by fiat.

### IV.B. Fiat Currency

Today when people talk about fiat currency they often mean currency created without being backed by real goods. However, the actual meaning of fiat currency is that it is any form of currency that comes about by edict rather than by the voluntary exchanges and contracts made within the market. All national currencies that have existed in recorded history are therefore fiat currencies, including commodity-backed currencies. Currencies that are backed by legal tender laws and are used for the payment of taxes in the country of origin will be defined as strong fiat currencies. Strong fiat currencies include the current national currencies used around the world and the previously used gold or silver standards. Weak fiat currencies are those that are chosen by edict to be used for trade or exchanges but are not backed by legal tender laws or strongly enforced. The best example of a weak fiat currency is the world's reserve currency.

This viewpoint is different from those who favor the idea of commodity-backed currency—namely currency backed by a finite resource such as gold or silver. These people typically advocate a return to a gold standard as a solution for the instabilities and imbalances in the world's economy. However, the real problem is that modern currency is a monopoly created by decree, a problem that a gold standard does not resolve. Additionally, the free market is supposed to utilize resources efficiently. Surely, the most efficient use of gold or silver cannot be to remain sitting in vaults.

The issue with fiat currency is simple. It does not arise from economic activity and is necessarily created by a subset of society. Even when other currencies may legally be allowed to operate within the economy, they cannot compete with the fiat currency because legal tender laws and the use of the fiat currency for payment of taxes give it a huge advantage over any competing currency. When a subset of society is granted a monopoly over the creation of currency, and the remainder of the society is required—to varying degrees depending on enforcement—to accept that currency in the settlement of debts, an imbalance is created in the economy that transfers wealth in the direction of the currency creators.

When a subset of society has a monopoly on currency creation, the question of how the other members of society get that currency to conduct their economic activity arises. Some might say that the responsibility of creating the currency is delegated to the government, which represents the people, and so it is acceptable for the government to create currency on behalf of the people. However, the government is also an actor within the economy and the question of how the remainder of society gets that currency still holds. Obviously the currency created cannot simply be handed over to the remainder of society as needed and in exchange for nothing, since in that case they might as well just create it themselves.

There are two ways by which the remainder of society can obtain the currency from the currency creators. They must either provide goods or services to the currency creators in exchange for the currency, or they must take loans from the currency creators and pay them back with interest. The interest payments become a profit for the currency creators who can then use those profits to purchase goods and services from the rest of society. While the former scenario is usually not allowed, it is worth noting that in today's economy, profits from trading in debt can be used to purchase goods and services. Thus the currency creators can trade in financial products based on the fiat currency that they create, and use the profits to purchase real goods and services.

This of course creates a problem of wealth transfer from the non-currency creators to the creators. However, that is not the only problem. Even within the remainder of society, which does not create currency, those who get first access to the fiat currency have a chance to spend it and use it to bid up prices of goods and services before others get that currency. There is thus a gradient that causes wealth to be transferred from those that get the currency last, usually the lowest classes, through those that get the currency soonest, usually the highest classes, to the banking and financial sector. The wealth transfer can also be in time as well as space. Older generations have access to the currency before newer generations. If there is inflation, i.e., the currency in circulation grows faster than the underlying economy, then people are able to consume more of society's savings than they otherwise would have been able to. Each successive generation therefore has access to less real savings than their preceding generation.

Fiat currencies can be in the form of fiat money, or fiat credit or debt. When backed by a real commodity, as in a gold standard, it is fiat money. When created without the backing of real goods or services as is done today, it can be considered as fiat debt. Fiat credit or debt cannot be considered as media of exchange because neither arises from exchange of goods and services in the economy. However together with fiat money, they can be considered media of payment even though they only gain that property by edict.

## V. SAVINGS

We have talked about the imbalance created by fiat currencies in the form of wealth transfer. However there are also other instabilities caused by the current currency system. To understand these instabilities we must look at the role of savings in an economy. The key factor for economic growth in a free market is savings. Savings represent deferred consumption and act as a hedge against risks. Risks include natural disasters, accidents, and economic risks such as the risk of starting a new business venture. The real savings of a society refers to the goods that have been produced but have not yet been consumed. The value of these savings is that they allow the members of the society to invest their time in economic activities other than those that satisfy their immediate needs. They can for example invest some of the time bought by their savings in people, to create new products, which result in economic growth. Without the time bought by savings, people in the society can only afford to spend their time producing goods that will be consumed immediately and cannot take the risk of investing their time in innovation.

Consider for example the case of the hunter-gatherer society of early humanity. If they had consumed all the food they hunted or gathered immediately, whether due to scarcity or an insatiable appetite, then they may not have evolved into an advanced society as we have now. If they consumed everything they came by, then they would have had no savings. With no savings they would have had no time for leisure or for taking risks and producing capital. Fortunately there was an abundance of food, and the animals hunted were large enough that they could afford to save some food for later. This would have meant that they did not have to use all their time for hunting and gathering but could afford some time for leisure, and those willing could start to take risks on producing tools, bringing about increased efficiency in the hunting, gathering and preparation of food.

The fact that savings is the hedge against risk explains the boom and bust cycle. The individual members of a society measure their savings—which is their share of the entire society's real savings—in terms of the amount of currency they hold or have access to. When central banks lower the interest rates and increase the amount of currency in circulation, or banks expand credit beyond that required by the prevailing economic activity, false signals are sent that there are more real savings—in goods and services—than there really are. These false signals cause people's time preferences to change and they start to consume more now instead of later because of this deception. What actually happens then is that while they falsely think they have a good hedge because they hold more currency, they are consuming more of their savings, hence reducing the hedge that society has against risks. This new consumption is seen as increased demand by businesses, which respond by investing more into current products or even new ones, creating a boom in the economy. A bubble may arise in a particular sector of the economy if most of the new consumption is concentrated in that sector. However, the investments made by businesses in this situation are malinvestments based on a false and unsustainable demand. As the real savings shrink, people begin to recognize this and demand falls, leading to a recession as the market tries to repair the damage caused by the boom and rebuild the savings. It is important to note that when you have a reserve currency system, the country that issues the reserve currency has access to the real savings of the rest of the world. Until the savings of the rest of the world can no longer support the reserve currency country, it can restart a boom by stimulus since that would draw savings from the rest of the world into the reserve currency country.

## VI. THE FREE MARKET

The contention of this paper is that the problems with our current currency system stems from the monopoly that government has on the creation of currency, and that all fiat currencies are inherently unstable. While some see it as a central banking problem, this paper argues that it is first and foremost a government monopoly problem with taxation being the core issue. The fact that governments are granted the power to tax the citizenry is the main problem because that is the source of the monopoly. If our currency system arose through a process other than banking, you would still have the same kind of monopoly problems only in a context different from banking. To fix the problem we must eliminate taxation completely.

In a free market, you would have a variety of competing currencies some of which would be classified as media of payment and others as media of exchange. Hayek (1990) previously considered the idea of competing currencies but his model included fiat currencies created by the banks. It should be noted that the currencies that exist in the free market do not necessarily have to be created by banks. Any person, or group of persons such as a corporation, can choose to issue a medium of payment or exchange and people are free to choose whether or not they will accept it as currency. This paper is mostly concerned with currencies that are considered media of exchange, i.e., they take the form of IOUs or invoices. The term free market is defined as the process by which people exchange goods and services voluntarily under rules of contract, meaning that one must honor any implicit or explicit contracts made.

As described by Mises (1998), humans act to reduce some uneasiness, meaning that they have a preference for being in a state that is different from the one they are currently in. In economic action the uneasiness is related to their assets versus their liabilities. An objective value—which is a function of time—is assumed to exist for the assets and liabilities held by the actor. The objective value of any good or service only exists relative to other goods or services in the form of the exchange ratio. The objective values are not known to the actors in the economy, who instead assign a subjective value to the various goods and services. Economic action will be defined as any action that is done to increase the subjective value of the actor's equity. This means any action that increases the subjective value of one's assets or decreases the subjective cost of one's liabilities. Assets include any property, knowledge or skills that an individual owns or has, and liabilities include any property or services owed to other people. There is also a cost to moving from one state to another. As before, the actor does not know the objective cost but instead assigns a subjective cost to the state change. The difference in subjective value between the current state and the preferred state is a measure of interest. If the interest exceeds the subjective cost of the state change, then an individual may act to change his state to the preferred one. The most efficient actors in the economy are those that make the best estimates of the values of assets and liabilities and the cost of changing from one state to another.

Exchange of goods and services in a society arises because people assign different subjective values to products. Thus two or more people may choose to exchange goods or services because the values they assign to the goods or services exchanged allows all parties to increase the subjective value of their equity through the exchange. This kind of exchange is completely voluntary and is the distinguishing characteristic of the free market.

We have previously looked at how money, or more specifically commodity money, arises though the free exchange of goods and services. We can also imagine how credit and debt, in the form of IOUs and invoices, can emerge as currency through the free exchange of goods and services. The use of IOUs and invoices as currency did not evolve naturally from barter because they required the invention of accounting first. We will thus imagine a free market where accounting was invented first and show how IOUs and invoices become currency.

**VI.A. IOUs As Currency**

Suppose two parties, *A* and *B* exchange goods or services. As stated previously, the exchange is in two parts. Say in the first half of the exchange, *A* provides a good of a certain value to *B*. If *B* does not immediately provide *A* with a good or service of the same or greater value—from the perspective of *A*—then *B* may issue an IOU to *A*. The IOU is a claim to future goods or services to be produced by *B*. If *B* has great credibility within the society, then a third party or multiple third parties may accept the IOU issued by *B* to *A* as payment for goods and services they provide. Thus IOUs issued by *B* can be denominated and become currency. When any such IOU is redeemed, the lifetime of that IOU ends. IOUs can therefore be used for indirect exchange in the same way as money. The nature of IOUs is that their value is dependent on how much faith people have in the party who issued it to honor the contract. The acceptance of IOUs from a particular issuer as payment is entirely voluntary.

In a free market economy, IOU currency from different issuers would exist within the economy. An example in the case of gold would be a gold mining company funding its operations using IOU currency. People may choose to accept IOU currency from a gold mining company because the proven reserves of its mines assures them that they can ultimately redeem the IOU currency for real gold. It is worth noting the difference between this and a gold standard currency system where banks issue certificates for gold bullion held in reserve. In the case of the banks, they are supposed to be holding gold that belongs to their customers, but in the case of the gold mining company the gold does not belong to the holders of the IOUs. Another example might be a large retail company that sells a wide variety of goods, funding all or part its operations in the same way by issuing IOU currency. People may choose to accept the IOUs because they can easily redeem them for goods that they use regularly in their lives.

**VI.B. Invoices As Currency**

It is easy to see how IOUs become currency. On the other hand it is not obvious how an invoice becomes currency, and so we generally don't think of invoices as currency even though they have a very important role to play in an economy. As people exchange goods and services in the market, there is a coincidence of their self-interests, which we call their mutual interests. These mutual interests arise from their need to make the market process more efficient. An example of this is the need for roads, which is a collective interest of a society arising from the individual interests of its members. Thus, in a free market economy, mutual interests create a demand for public goods and services.

When economists have looked at the efficiencies of free markets previously, they have only looked at its efficiency in satisfying the individual self-interests of the actors within the economy. When it comes to public goods and services, it has been taken for granted that it is the role of government to provide and/or manage the provision of these goods and services, and that government has the right to determine how much each person will contribute as payment for the provision of these public goods and services. However, as has been pointed out, public goods and services are just goods and services that people are jointly interested in having. Thus, they still remain in the self-interest of the actors in the economy, although only to the extent that they are engaging in economic interaction with other members of the society. A true free market, therefore, should not only efficiently provide goods and services to meet the private needs and wants of the actors, but should also efficiently provide public goods and services. A true free market would be expected to find a balance between the provision of private and public goods and services.

The problem with public goods and services is that it is difficult to determine how much each member of society should contribute to pay for the goods or services. The way it has been achieved since the advent of modern civilization has been through taxation, whereby government is granted a monopoly over the provision of public goods and services, the authority to determine what percentage of each person's income they must contribute towards these public services, and the right to confiscate a part of each person's income in order to pay for the public goods and services that government provides.

Taxation goes against the principles of the free market since it is not entirely voluntary, and has several limitations with regards to efficiency. It requires the commitment of resources and labor to the collection of taxes, does not truly account for how much each individual actually uses the public goods or services, and also results in individuals committing time and energy to minimizing the amount of taxes they have to pay. The idea that taxation is necessary is based on an implicit belief that everyone must contribute directly to the provider of the public good or service. However, this is not true for the same reason that individuals do not have to engage only in direct exchange. Indirect exchange is much more efficient as is well known. It is only necessary that the provider of the public good or service be compensated. The concept of invoices as a form of currency allows for the compensation of providers of public goods and services in a way that is fair and efficient in the manner of indirect exchange.

In order to understand how invoices become currency, it is important to look at the exchange in its complete context. With IOUs, even though they are initially issued by one party to another, they become currency because other members of the society voluntarily choose to accept the IOUs as a medium of payment. Thus, it is within the context of the entire society that IOUs become currency. The same is true for invoices except that the concept is inverted. Here the party that receives the good or service is a group of people within the society, or even the entire society itself. The invoice is issued by the provider of the public service—in the first half of the exchange—or by a party delegated to do so on his behalf. The invoice can automatically be denominated and used as a currency by the recipients of the public service because they are bound by contract to accept this invoice, having received the public good or service.

## VII. PUBLIC SERVICE INVOICES

Contrary to the long held view that taxation is a necessary part of civilization, the concept of invoices as currency shows that it is not necessary to tax the members of a society directly in order to fund the provision of public goods and services. "Taxation" by indirect exchange, which is much more efficient, is possible with the introduction of invoices as currency. When invoices are issued for goods or services provided to a group of people, the invoice can become a currency for that group of people. In particular, invoices issued for the provision of public services to a society can become a currency for that society, and as they use that currency, it automatically implements a "taxation" by indirect exchange and eliminates the need for taxes to be collected in any form. An invoice issued for a public service will be referred to as a Public Service Invoice, or PSI for short. How PSIs work is illustrated with the following example:

> Suppose you have a society going about its normal business. They may use barter or have a number of private currencies but they don't have a currency acceptable to all members of the society. They wish to build a road network to facilitate their economic activity, so they contract someone to do it for them. Now the way we do things today is we will take contributions—in the form of currency—from everyone in the community engaging in economic activity, i.e., tax them and use that to pay the contractor, but that is inefficient. A more effective way is for the contractor, upon providing the roads, to create a PSI, which he then needs to give to the society. That PSI can be denominated into a currency, which the people in the society must accept because it is an invoice for a service they requested and received. This is exactly the same as an individual accepting an invoice from a service provider, except in this case the invoice can be denominated into a currency because the recipient of the service is a group of people. Using the PSIs he has created in the form of a currency, the contractor can then purchase goods and services or even private currencies from others in the society and he is thus compensated for his services.
>
> What happens to the PSI currency as it circulates through the society? Suppose for example the contractor spent all the PSIs he created in one shop. That shop owner would then have all the PSIs and would have provided to the contractor, goods or services equal to, or more than, the value of the roads provided—from the viewpoint of the contractor. At this initial stage that shop owner has basically paid all the "taxes" for the roads. Note that having paid all the "taxes", he holds all the PSIs. As the shop owner spends the PSIs within the society to restock his shop, he will be paid —in goods and services—by those he buys from. So his suppliers and he would now have PSIs representing how much in goods and services they have provided directly or indirectly towards compensating the contractor. At this second stage, the total of the "taxes" would have been paid in part directly by the shop owner, and the remaining part, indirectly by his suppliers. The shop owner paid "taxes" in goods and services provided to the contractor while the suppliers paid "taxes" indirectly by providing goods and services to the shop owner. As the invoice currency is spent in the society it will spread amongst all participants in the economy, and whatever amount of that currency a person holds represents what he has contributed in goods and services towards the roads that were built.

This completely inverts the idea of taxation. Instead of people paying taxes by having a portion of their income taken by government, they pay their "taxes" by the actual goods and services they provide to others in the society. By providing a good or service in exchange for PSIs, they directly or indirectly compensate the providers of public services. Instead of losing currency when they pay taxes, they gain PSIs instead. The amount of PSIs a person holds represents the net contribution—in goods and services—he or she has made towards the payment for public goods and services. Another way of thinking about it is that the PSIs are a reward for contributing some goods or services to society. That reward can then be used to claim some goods or services from society in the future.

Because invoice currency represents goods or services received in the past while IOU currency represents goods or services to be received in the future, the invoice currencies will always be spent preferentially over the IOU currencies. Clearly if you have to pay someone and the person would either accept the invoice currency or an IOU currency, you would rather spend the invoice currency. As a result of this, the PSIs will distribute very quickly within the society and come to a fairly stable distribution representing the economic activity of the individuals in the society. However, PSIs are also fluid and if the economic activity of the actors changes, they will quickly redistribute as people spend, to reflect the change in activity.

It is easy to see that whoever provides the most value to the society with his goods or services will also have the most PSIs. This is the correct way that taxation should work—the more value you contribute to others in the society, the more you are rewarded with increased purchasing power. Note that in a true free market the government and the banking and financial sector would not have the special privileges that give them a monopoly in currency creation, so they will not have an advantage over other sectors of the economy. They would not be rewarded by the market for their role as currency creators, because no one is forced to accept that currency anymore. The new system inverts the current idea of taxation to a more correct form. Instead of people being asked to contribute more to pay for public services because they make the most in terms of earning money and credit, we see that people earn more PSIs as they contribute more—directly or indirectly in terms of goods and services—towards the compensation of public service providers. If the amount of currency held by a person is what would be considered a measure of wealth, then the most wealthy would be those who contribute the most in goods and services to the remainder of society.

In the currency model described above, the value of PSIs would be determined by the supply and demand of public services. Public service providers have to provide a service first, before PSIs can be issued. Thus, the service providers would have to raise funds in the private sector for their public projects, but they have an assurance of payment—in real goods and services—from the public who must accept the PSIs that will be issued for those services they requested. Such a system would therefore bring market forces into the domain of public services where they had previously not been a significant factor. This currency model also allows people to determine how much in public services they want or can afford, reflected in the value of goods and services they are willing to give up to the public service providers. The balance between private and public goods and services is found by the use of the two kinds of currencies, namely IOUs and invoices. Private IOU currency would primarily be used for saving and investment, while invoice currency would typically be used for everyday spending. If there is a huge demand for public services and the amount of invoices in circulation increases, then more IOU currency is saved and invested. Conversely, if there is only a small demand for public services and therefore a limited supply of PSIs, more IOU currency is spent so there is no shortage of currency for everyday consumption.

The lifetime of any PSI unit ends when it is returned to the issuer. Since governments provide services at both the public and individual level, they can issue PSIs for the public projects and services while accepting them as payment for individual services. When a PSI unit returns to the government it is destroyed and can no longer go back into circulation. Thus PSIs are simply a method of accounting for how much each person has contributed in goods and services to the remainder of society.

It is important to note that PSIs must specify the public good or service provided, just as IOUs specify the good or service they can be redeemed for. Thus, individuals have the opportunity to confirm that any PSI unit offered to them as compensation represents a good or service that was requested from the government. If it does not, then they have the right to reject it. A responsibility is therefore placed on the public service providers themselves to make sure that any service they provide was actually requested by the people in the society, otherwise they may not be able to receive compensation for it.

It should also be noted that while the society must generally accept PSIs, they do so not because of legal tender laws but because they are bound by contract. PSIs are produced for goods or services that society requested and received through their representatives. They must accept them just as an individual must accept an invoice for goods or services received. However, like the individual, they have the right to refuse the PSI if it was issued for a service they did not request.

**VII.A. Political Implications**

Clearly there are political implications to having such a currency system. The first and most obvious is that it requires some changes to the structure of government. The Judiciary, previously thought of as an independent branch of government must now be independent of the government itself, since it must adjudicate on cases across all scales including disputes between the government and the people. Everyone, including the government, is considered an actor in the economy, and must operate under the rules of contract, which will be enforced by the independent judiciary. The executive branch also no longer has a monopoly on managing or providing public services, as now anyone can be a public service provider. Instead, a new branch of government may be created that manages the compensation of public service providers.

The competing currencies model, with private currencies operating alongside the PSIs, enables the populace to have the freedom to reject PSIs for any public project or service that was not requested by the people directly or through their representatives. They have the ability to do so because other currencies can compensate for any shortages that arise due to the rejection of those invoices. Thus, the suggested currency model equalizes power between government and the people by making the government accountable for its actions through the PSIs issued. We see that it is only without taxation or government currency monopoly that we can have a true government of the people.

## VIII. FUNDING OF GOVERNMENT

Governments today are funded in three ways: taxation, inflation and borrowing. The elimination of taxation has already been discussed, but inflation and borrowing must still be addressed. In the new currency model described above, PSIs are created to pay for public services, and the value of the PSIs in circulation is determined by the supply and demand for public services. PSIs are issued for actual goods and services provided to the public. Thus the growth of PSI units in the economy is not inflationary since it is backed by real economic growth. While there might be some price inflation in PSI currency—due to rapidly increasing supply of the currency—when it is initially adopted, in the medium to long term it should stabilize because there is only a finite supply of public goods and services over any given period and demand would not increase exponentially. In addition, there is a mechanism by which PSI units are withdrawn from circulation as individuals make use of government services.

There will also no longer be a need for government borrowing. While governments typically borrow money to fund their projects and services, the only reason for any country to borrow would be if they did not have the resources they need already in the country. It makes no sense for a country to have to borrow from a bank or other country just to direct resources it already has towards a project. With the use of PSIs, resources within the country can be efficiently directed towards public services and projects without borrowing. When the country needs additional resources, they can be brought into the country, and paid for, through the import and export process which replaces government borrowing, i.e., real goods and services are exchanged rather than borrowing in currency.

## IX. STABILITY

With the new currency system proposed, currency creation is no longer limited to a subset of the population. Any individual or group of individuals can choose to create a currency and their reputation will determine how widely accepted the currency becomes. This means that the gradient that transfers wealth in the direction of monopoly currency creators no longer exists. Note that even with the PSIs, anybody can become a public service provider so PSIs are not limited to being issued to a subset of society. Also the public service providers must provide a good or service to the society first, before receiving their PSIs, so currency is not being created without an exchange of goods and services taking place.

Additionally, banks no longer control the currency. There will be private currencies in circulation, in addition to the national PSIs. With no central bank, the market will set interest rates. Also since PSIs represent actual goods provided, banks will not be allowed to expand that currency using fractional reserve banking. Thus the main issues that cause instability in the currency will be removed for the PSIs. There might still be issues with the private currencies but these should be small and should not destabilize the entire economy.

## X. CONCLUSION

The currency system used by the world today is an artificial construct that evolved from the idea of banking. The problem with the current fiat currency system and the way it evolved is that it leads to a partnership between the government and the banking and financial sector, which acquire a monopoly in the creation of currency. When a subset of society has a monopoly on currency creation, a transfer of wealth occurs, as those in the remainder of the society become more and more indebted to those that create the currency. It is also impossible for this debt to be paid off because the amount of currency in circulation is always less than the debt owed. The result is an ever-widening gap between the rich and the poor. Other instabilities such as the boom and bust cycle occur, as the currency creators have no way of knowing exactly how much currency they need to create.

This paper shows that voluntary exchange and indirect exchange can take place at all scales within an economy. Just as the concept of IOUs as currency arises at the individual level, a concept of invoices as currency arises on the scale of an entire society. The two kinds of currencies together in circulation provide a balance between individual needs and wants, and the needs and wants of the society as a collective. The efficiency of indirect exchange at the individual level is also found at the societal level and the need for direct taxation of the citizenry is eliminated. The new currency system resolves the problems with the fiat currency system because currency creation is not limited to a subset of society, and there are currencies created through processes other than debt, ensuring that there is enough currency for all interest payments to be made. PSIs provide a way to have a national currency without the need for legal tender laws. Common currencies in a true free market would emerge from voluntary exchanges between the public and public service providers. With the use of PSIs as described in this paper, there is still the inefficiency of using the legislative process to request public goods and services. However, finding a more efficient method for requesting public goods and services would require a more general exercise and is beyond the scope of this paper.

In a society where people's interests are varied, it is highly unlikely that their goals are sufficiently aligned so that the society can be managed as with a household or with a company. The attempt to manage the economic affairs of a nation or state can result in instabilities due to unintended consequences. For example, in attempting to create a so called wealth effect, central banks deceive the people in the society into consuming their savings and hence decreasing the hedge that society has against risks, resulting in the boom and bust cycle. By moving to a currency system that is not centrally managed, we get a system that is more able to account for people's time preferences and to quickly respond to changes in these preferences. This would take us one step closer to having a true free market.